\documentclass[aps,twocolumn,color,psfig,epsf]{revtex4}
\usepackage{amsmath}
\usepackage{color}
\usepackage{amsfonts}
\usepackage{epsf}
\usepackage{graphicx}
\usepackage{epstopdf}
\usepackage{ulem}

\usepackage[position=top,caption=false]{subfig}

\newcommand{\rr}{{\mathbf r}}

\begin{document}

\title{Debye-H\"uckel potential at an interface between two media}

\author{Alexander Morozov, Iain Muntz, Job H.~J. Thijssen and Davide Marenduzzo}
\affiliation{SUPA, School of Physics and Astronomy, The University of Edinburgh, Edinburgh, EH9 3FD, Scotland, United Kingdom}

\begin{abstract}
Electrostatic interactions between point charges embedded into interfaces separating dielectric media are omnipresent in soft matter systems and often control their stability. Such interactions are typically complicated and do not resemble their bulk counterparts. For instance, the electrostatic potential of a point charge at an air-water interface falls off as $r^{-3}$, where $r$ is the distance from the charge, exhibiting a dipolar behaviour. This behaviour is often assumed to be generic, and is widely referred to when interpreting experimental results. Here we explicitly calculate the in-plane potential of a point charge at an interface between two electrolyte solutions with different dielectric permittivities and Debye screening lengths. We show that the asymptotic behaviour of this potential is neither a dipole, which characterises the potential at air-water interfaces, nor a screened monopole, which describes the bulk behaviour in a single electrolyte solution. By considering the same problem in arbitrary dimensions, we find that the physics behind this difference can be traced to the asymmetric propagation of the interaction in the two media. Our results are relevant, for instance, to understand the physics of charged colloidal particles trapped at oil-water interfaces.
\end{abstract}

\maketitle

The physics of charged objects at, or near, liquid interfaces is full of subtle and non-trivial effects~\cite{Stillinger1961,Pieranski1980,Netz1999,Jancovici1980,Attard1988,Leunissen2007,Naji2014,Vermant2010,Aveyard2002,Frydel2007}. For instance, charged particles trapped at the air-water interface interact electrostatically via a long-range dipole-dipole repulsion, even if mobile ions screen any Coulombic interaction in the water phase~\cite{Pieranski1980,Hurd1985}. An oil-water interface can spontaneously acquire a non-zero charge due to the adsorption of hydroxyl ions~\cite{Marinova1996}. This surface charge leads to a generic repulsion between uncharged colloidal hard spheres trapped on such an interface~\cite{Muntz2019}. 
Electrostatics at interfaces is also important in soft matter and biological physics, as it underlies the behaviour of proteins moving on a lipid bilayer, or the adsorption of DNA or other polyelectrolytes on charged membranes~\cite{Netz1999,Slosar2006}. 

A popular approximate treatment for electrostatic effects in a bulk electrolyte is the Debye-H\"uckel theory~\cite{Debye1923}, which is valid when the electrostatic potential is small everywhere in the system so that non-linear effects can be disregarded. Whilst simplified, this theory can be formulated so as to include the effects of ionic fluctuations at a Gaussian level~\cite{Netz1999}. The main successful prediction of the Debye-H\"uckel theory is that mobile ions in an electrolyte generically screen charges, so that the potential of a point charge is proportional to $e^{-\kappa r}/r$, instead of being $\sim 1/r$. The quantity $\kappa$ is the inverse of the Debye screening length: it depends on ionic charge and concentration, and quantifies the efficiency of screening. 

\begin{figure}[!h]
\centerline{\includegraphics[width=0.4\textwidth]{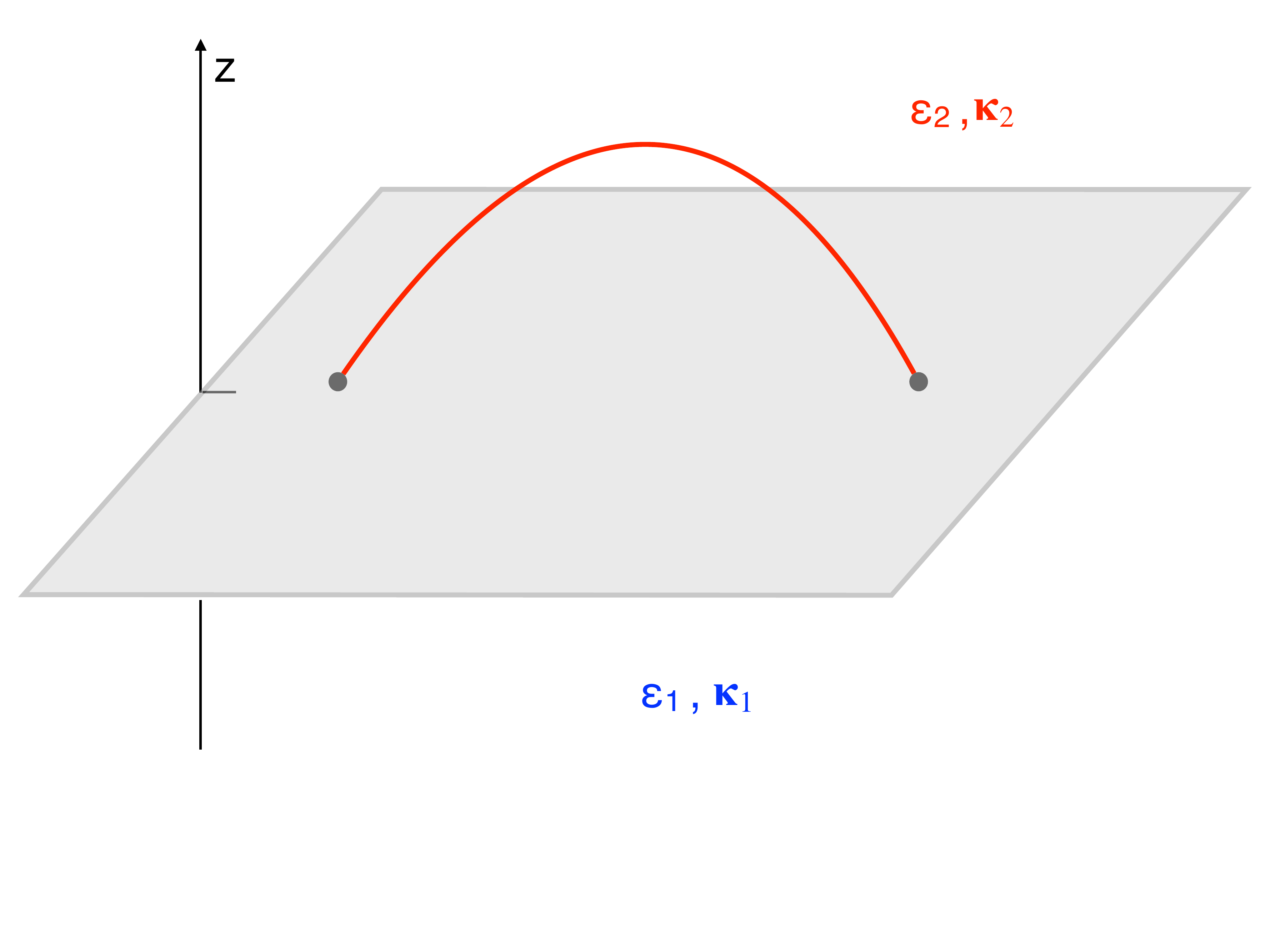}}
\caption{Schematics of the problem we consider. A pair of charged point particles lies at the interface between two electrolytes with different screening length and dielectric permittivities. The red line connecting the two particles is an example of a Debye string contributing to the calculation of the interparticle potential (see text).} 
\label{fig0}
\end{figure}

The Debye-H\"uckel theory has been generalised for systems with an interface~\cite{Stillinger1961,Pieranski1980,Netz1999,Hurd1985}, with most results focussing on the case where one of the electrolytes has no mobile ions, so that its Debye length is infinite. In this case, which is directly relevant to air-water interfaces, the interaction potential of a point charge along the interface was shown to decay as $r^{-3}$~\cite{Hurd1985}, with the decay law resembling bulk dipolar interactions. While this result is not directly relevant to interfaces separating media with finite, yet different screening lengths, like oil-water interfaces, it is often assumed that there should at least be a large range of distances over which the $r^{-3}$-decay is observed in such situations as well. Here we demonstrate that this is in general not the case. We study the simple but fundamental problem of a point charge at an interface between two electrolytes with different (but finite) Debye screening lengths (Fig.~\ref{fig0}). We show that the in-plane potential at the interface decays with an anomalous scaling, which differs from both the screened Coulomb potential characterising charge interactions in bulk electrolytes and the dipolar decay, relevant for water-air interfaces. We argue that the potential we derive should determine the physics of self-assembly in colloidal monolayers at oil-water interfaces, such as those which are formed, for instance, in ``bijels''~\cite{Cates2008} or Pickering emulsions~\cite{Pickering1907,Ramsden1903}.

The problem we are interested in is sketched in Fig.~\ref{fig0}. Two point particles, each with charge $Q$, lie at the interface between two dielectric media, with dielectric permittivities $\epsilon_1$ and $\epsilon_2$ respectively. The interface is normal to the $z$ axis and located at $z=0$, whereas $\rr$ denotes positions on the plane parallel to it (Fig.~\ref{fig0}). While there are mobile ions in each of the two media, there are no ions at the interface (although the case of a salty interface with mobile ions can be dealt with~\cite{Netz1999}, as mentioned in the conclusions). We want to find the interparticle potential $U(r)=Q\phi(r)$, where $r$ is interparticle distance and $\phi(r)$ is the value of the electrostatic potential generated by the first particle at the position of the second particle. The equation for $\phi$ is given by the Maxwell equation $\nabla \cdot {\mathbf D}=\rho$, where ${\mathbf D}$ is the electric displacement field and $\rho = Q\delta(\rr)\delta(z)+\rho_{\rm ion}(\rr,z)$ is the total charge density, which includes the interfacial point charge and the ionic charges in the two media, $\rho_{\rm ion}(\rr,z)$ (note $\delta(\rr)$ refers to a two-dimensional Dirac delta function).

Following~\cite{Stillinger1961}, we can find $\phi$ by using the Poisson-Boltzmann theory in each of the two media to approximate $\rho_{\rm ion}=n_0 e^{-\frac{Z e_0\phi}{k_BT}}$, with $n_0$ the ionic concentration in the bulk of that medium, $Z$ the valence of the ions, $e_0$ the elementary charge, $k_B$ the Boltzmann constant, and $T$ the temperature. The linearised version of this equation, valid for $\frac{Z e_0\phi}{k_BT}\ll 1$, is given by
\begin{equation}\label{interfaceDebyeHueckel}
\nabla\cdot \left(\epsilon(z)\nabla \phi\right) - \epsilon(z)\kappa^2(z) \phi = -Q \delta(\rr)\delta(z)
\end{equation}
where $\kappa(z)$ is the inverse Debye length and $\epsilon(z)$ the dielectric permittivity of the medium. For our geometry, the parameters ($\epsilon(z)$,$\kappa(z)$) equal ($\epsilon_1$,$\kappa_1$) for the first medium ($z<0$), and ($\epsilon_2$,$\kappa_2$) for the second medium ($z>0$). For an oil-water interface, such as dodecane-water, typical values are $\kappa_1\sim 10\kappa_2 \sim 1$ $\mu$m$^{-1}$, and $\epsilon_1\sim 40 \epsilon_2\sim 80 \epsilon_0$ (with $\epsilon_0$ the dielectric permittivity of free space).

Introducing the in-plane Fourier transform, so that $\phi(\rr,z)=(2\pi)^{-2} \int d{\mathbf q} e^{i{\mathbf q}\cdot\rr} \hat{\phi}({\mathbf q},z)$, in Eq.~(\ref{interfaceDebyeHueckel}), we obtain the electrostatic potential in the two half-spaces,
\begin{eqnarray}\label{potential1}
\hat{\phi}({\mathbf q},z) & = & A e^{\sqrt{q^2+\kappa_1^2}z} \qquad z<0\\ \nonumber
\hat{\phi}({\mathbf q},z) & = & B e^{-\sqrt{q^2+\kappa_2^2}z} \qquad z>0.
\end{eqnarray}
From Eq.~(\ref{interfaceDebyeHueckel}), it can be seen that the potential needs to be continuous at $z=0$, so that $A=B$, and that there needs to be a discontinuity in its derivative, such that,
\begin{equation}\label{derivativeBC}
\epsilon_2 \left[\frac{\partial \phi}{\partial z}\right]_{z\to 0^+}-\epsilon_1 \left[\frac{\partial \phi}{\partial z}\right]_{z\to 0^-}= -Q.
\end{equation}

As a result, the potential as a function of position on the interface is given by the following ($2$-dimensional) inverse Fourier transform~\cite{Stillinger1961,Hurd1985,Netz1999},
\begin{eqnarray}\label{potential}
\phi(r) & = & \frac{Q}{4\pi^2} \int d^2 {\mathbf q} \frac{e^{i {\mathbf q}\cdot \rr}}{\epsilon_1 \sqrt{\kappa_1^2 + q^2} + \epsilon_2 \sqrt{\kappa_2^2 + q^2} } \\ \nonumber 
& = &\frac{Q}{2\pi r} I(r),
\end{eqnarray}
where
\begin{align}
I(r) = \int_{0}^\infty dx \frac{x J_0(x)}{\epsilon_1 \sqrt{\kappa_1^2 r^2 + x^2} + \epsilon_2 \sqrt{\kappa_2^2 r^2 + x^2} },
\end{align}
$r=|\rr|$, and we have introduced the zero-th order Bessel function of the first kind, $J_0$. The asymptotic behaviour of the interaction potential between two interfacial point charges, $U(r)=Q \phi(r)=Q^2 I(r)/(2\pi r)$ is determined by the integral $I(r)$, which we study below.

Tailoring the procedure in~\cite{Hurd1985} to our system, we express the integral $I(r)$ as
\begin{align}
I(r) = \frac{1}{\epsilon_1^2 - \epsilon_2^2} \Bigg[ \epsilon_1 I_1(r) - \epsilon_2 I_2(r) \Bigg],
\end{align}
where
\begin{align}
I_i(r) =  \int_{0}^\infty dx \frac{x J_0(x) \sqrt{\kappa_i^2 r^2 + x^2}}{\alpha r^2 + x^2}, \quad i=1,2,
\end{align}
with 
$\alpha = (\epsilon_1^2 \kappa_1^2 - \epsilon_2^2 \kappa_2^2  )/(\epsilon_1^2 - \epsilon_2^2)$.
We now introduce $\delta_i \equiv \alpha - \kappa_i^2$, 
and Taylor-expand $I_i(r)$ with respect to $\delta_i$, to obtain the following series of integrals,
\begin{equation}
I_i = \sum_{m=0}^\infty (-1)^m (\delta_i r^2)^m  \int_{0}^\infty dx \frac{x J_0(x) }{\left(\kappa_i^2 r^2 + x^2\right)^{m+1/2}}.
\end{equation}

As detailed in the Appendix, these integrals can be computed exactly to give, as a final result,
\begin{widetext}
\begin{align}
I_i = e^{-\kappa_i r} - e^{-\kappa_i r} \frac{r\delta_i}{\kappa_i} \sum_{p=0}^{\infty} \frac{(-1)^p}{2p+1} \left( \frac{\delta_i}{\kappa_i^2}\right)^p 
{}_1F_1\left(2p+1;\frac{3}{2}+p;-\frac{r\delta_i}{2\kappa_i} \right),
\end{align}
\end{widetext}
where ${}_1F_1$ is the confluent hypergeometric function of the first kind~\cite{Abramowitz1972}. The asymptotic behaviour for large values of $r\delta_i/(2\kappa_i)$ is given by~\cite{Abramowitz1972} (see Appendix)
\begin{equation}\label{asymptotic0}
I_i \sim -e^{-\kappa_i r}\left[\frac{\kappa_i}{r\delta_i}+\frac{3\kappa_i^2}{r^2\delta_i^2}+\frac{1}{r^2\delta_i}+{\mathcal O}\left(\frac{1}{r^3}\right)\right].
\end{equation}

For an oil-water interface, the integral $I$ is dominated by 
$I_2$ for $r\to\infty$ (as $\kappa_2<\kappa_1$). 
The corresponding leading asymptotic behaviour for the interaction potential $U$ between two interfacial point charges is
\begin{align}\label{asymptotic2}
U(r) \sim \frac{Q^2}{2\pi}\frac{\epsilon_2\kappa_2}{\epsilon_1^2(\kappa_1^2-\kappa_2^2)}\frac{e^{-\kappa_2r}}{r^2}.
\end{align}
As $\kappa_2\to 0$, which would be relevant for an air-water interface, Eq.~(\ref{asymptotic2}) vanishes, so that we need to take the next term in the expansion, and we recover the dipole contribution, $U(r)\sim 1/r^3$, previously found and discussed in~\cite{Pieranski1980,Hurd1985}. Notably, however, if $\kappa_2\ne 0$, Eq.~(\ref{asymptotic2}) differs from, and decays faster than, a screened monopole with decay constant $\kappa_2$: we obtain $U(r) \sim e^{-\kappa_2 r}/r^2$, rather than $\sim e^{-\kappa_2 r}/r$. As expected, a simple screened monopole behaviour is found, from Eq.~(\ref{potential}), in the limiting case in which $\kappa_1=\kappa_2$, where there is no interface. 

The exact functional form of $U(r)$, obtained by numerically evaluating the integral in Eq.~(\ref{potential}), is compared to the asymptotic behaviour coming from Eq.~(\ref{asymptotic0}) in Fig. 2, for an oil-water interface. A good fit to the numerical solution for all $r$ is provided by
\begin{equation}\label{fit}
U(r) \simeq \frac{Q^2}{2\pi \epsilon_1} \left[\frac{e^{-\kappa_1 r}}{r} + \frac{\epsilon_2}{\epsilon_1 \kappa_1^2} \frac{e^{-\kappa_2 r}}{r^3} +\frac{\epsilon_2\kappa_2}{\epsilon_1 \kappa_1^2} \frac{e^{-\kappa_2 r}}{r^2} \right]
\end{equation}
where we have accounted for the fact that $\kappa_2\ll \kappa_1$ for an oil-water interface. Eq.~(\ref{fit}) describes a crossover between a screened monopole-like behaviour at small $r$ -- with the decay length equal to that of the first medium, $\kappa_1$ -- and the $\sim e^{-\kappa_2 r}/r^2$ behaviour at large $r$. At intermediate $r$ the second term, which is a dipole-like contribution (with screening, as $\kappa_2\ne 0$), can in principle play a role. For an oil-water interface the crossover between screened monopole and dipole is at $r_{c,1}=-\frac{2}{\kappa_1-\kappa_2}W_{-1}\left(-\frac{1}{2}\sqrt{\frac{\epsilon_2}{\epsilon_1}}\frac{\kappa_1-\kappa_2}{\kappa_1}\right)$ where $W_{-1}$ denotes the negative branch of the Lambert function~\cite{Abramowitz1972}; the crossover between dipole and asymptotic behaviour instead occurs at $r_{c,2}=\kappa_2^{-1}$~\cite{note}.
For parameters relevant to a dodecane-oil interface (Fig.~1), $r_{c,1}\simeq 9$ $\mu$m and $r_{c,2} \sim 10$ $\mu$m, so that the dipole regime is essentially absent. For water-oil interfaces with $\epsilon_1/\epsilon_2\sim 40-100$, the screened dipole regime is of practical relevance only if $\kappa_1/\kappa_2\gg 10$ (see Fig.~\ref{fig1b} and Appendix).


\begin{figure}[!h]
\centerline{\includegraphics[width=0.5\textwidth]{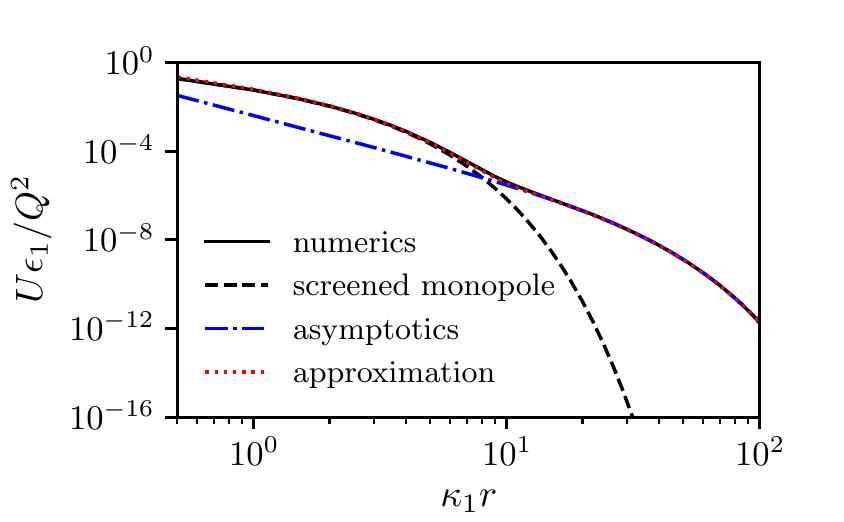}}
\caption{Log-log plot of the numerical solution of $U\epsilon_1/Q^2$ versus $\kappa_1 r$, showing the crossover between small $r$ behaviour, corresponding to a screened monopole with decay constant $\kappa_1$, and asymptotic behaviour, computed via Eq.~(\ref{asymptotic0}). The approximation in Eq.~(\ref{fit}) is also shown. Parameters are: $\kappa_1=10\kappa_2$, $\epsilon_1=40\epsilon_2$, relevant for a water-oil interface.}
\label{fig1}
\end{figure}

\begin{figure}[!h]
\centerline{\includegraphics[width=0.4\textwidth]{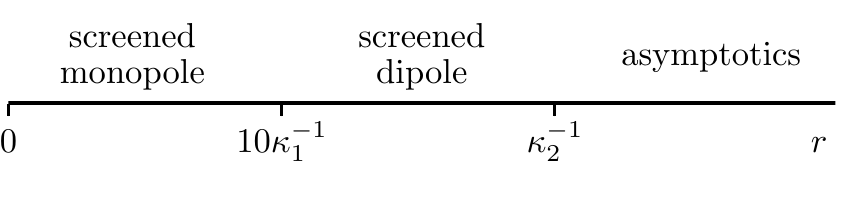}}
\caption{Diagram showing the ranges in $r$ where each of the regimes in Eq.~(\ref{fit}) dominates for the potential between two point charges at an oil-water interface with $\epsilon_1=80 \epsilon_2$.}
\label{fig1b}
\end{figure}

To gain more physical insight into the physics behind Eq.~(\ref{asymptotic2}), it is useful to consider the same interfacial Debye-H\"uckel problem defined by Eq.~(\ref{interfaceDebyeHueckel}) in arbitrary dimension, $d$. This problem is equivalent to that of finding the $d$-dimensional Yukawa interaction at an interface. 
We find that, for generic $d\ge 3$, the interfacial potential $\phi(r)$ is given by
\begin{eqnarray}\label{potentialddimensions}
\phi(r) & = & \frac{Q}{\left(2\pi\right)^{\frac{d-1}{2}} r^{d-2}} I_d \\ \nonumber
I_d & = &  \int_{0}^\infty dx \frac{x^{\frac{d-1}{2}} J_{\frac{d-3}{2}}(x)}{\epsilon_1 \sqrt{\kappa_1^2 r^2 + x^2} + \epsilon_2 \sqrt{\kappa_2^2 r^2 + x^2} },
\end{eqnarray}
where $J_{\alpha}(x)$, with $\alpha$ a real number, denotes the Bessel function of the first kind of order $\alpha$. 

\begin{figure*}
\centerline{\includegraphics[width=0.95\textwidth]{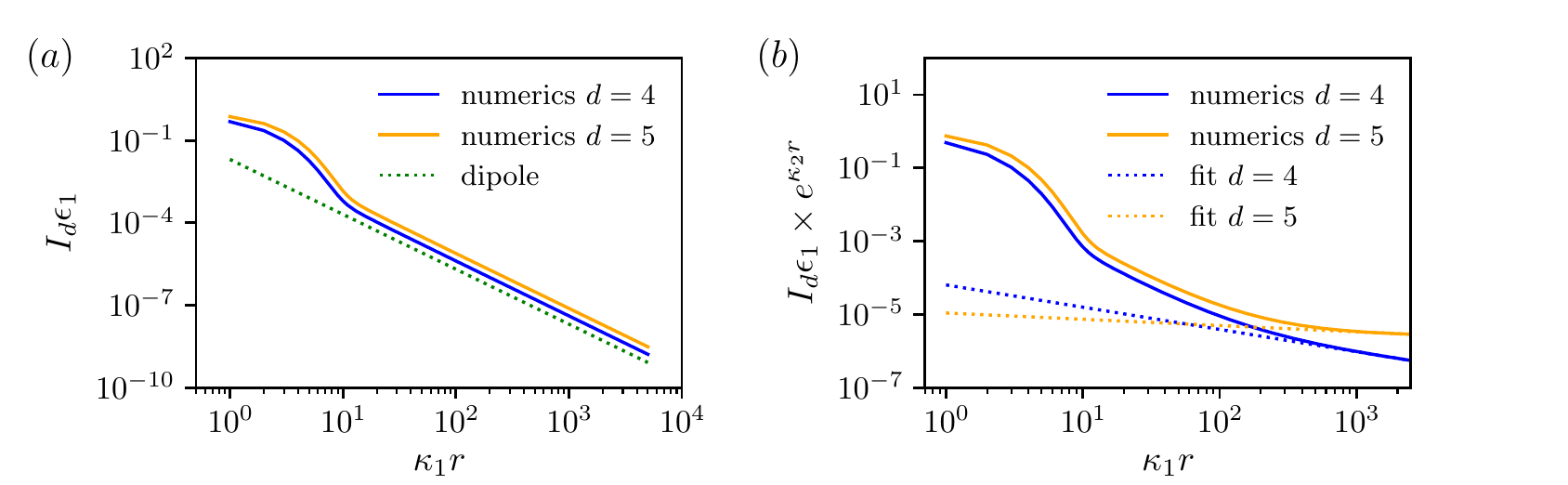}}
\caption{(a) Log-log plot of the numerical solution of ${I_d}\epsilon_1$ 
versus $\kappa_1 r$ for $d=4,5$, $\epsilon_1=40\epsilon_2$ and $\kappa_2=0$. The decay corresponding to a dipole-dipole interaction is $1/r^{2}$ for any $d$, which is plotted as a dotted line.
(b) Log-log plot of the value of $I_d\epsilon_1 e^{\kappa_2 r}$ versus $\kappa_1 r$ for $d=4, 5$, $\epsilon_1=40\epsilon_2$ and $\kappa_1=100\kappa_2$. The dotted line denotes fit to power laws at large distances (see text).}
\label{fig2}
\end{figure*}

Figure~\ref{fig2} shows the results from numerically evaluating the integral in Eq.~(\ref{potentialddimensions}), $I_d$, 
for $d=4, 5$. 
In the limit in which $\kappa_2=0$, which is the $d$-dimensional analogue of an air-water interface, we obtain that $I_d\sim r^{-2}$, independent of $d$. Consequently, $\phi(r)\sim 1/r^{d}$ (Fig.~\ref{fig2}a), which is the decay of an electrostatic dipole-dipole interaction in $d$ dimensions. This is consistent with the physical picture that if one of the media is free from ions, the interparticle interaction is a repulsion between the dipoles which emerge as counterions in the aqueous phase assemble close to the point charges
~\cite{Pieranski1980,Hurd1985}.

In the case when $\kappa_2\ne 0$ (and $\kappa_2 \ll \kappa_1$), which gives the $d$-dimensional analogue of an oil-water interface, we find that the asymptotic behaviour of $I_d$ instead depends on $d$. Fitting $I_d$ for large values of $r$ numerically, we find $I_4 \sim r^{-0.61}$ and $I_5 \sim r^{-0.17}$. These numerical behaviours are close to, and consistent with, $I_4 \sim r^{-0.5}$ and $I_5 \sim r^0$, or a constant. Therefore in generic $d\ge 3$ our numerical data are compatible with the asymptotic behaviour (see Appendix) 
\begin{equation}
\phi(r) \sim \frac{e^{-\kappa_2 r}}{r^{\frac{d+1}{2}}}.
\end{equation}
As in $d=3$, while the dominant contribution is screening with a typical length $\kappa_2^{-1}$, there is a different power law correction with respect to the Yukawa potential in the bulk, which is $\phi(r)\sim \frac{e^{-\kappa_2 r}}{r^{\frac{d-1}{2}}}$. In this $d$-dimensional case, a suitable approximation for the potential is therefore (see Appendix)
\begin{equation}\label{fitddimensions}
\phi(r) \simeq A_d \frac{e^{-\kappa_1 r}}{r^{\frac{d-1}{2}}} + B_d \frac{e^{-\kappa_2 r}}{r^{\frac{d+1}{2}}},
\end{equation}
where $A_d$ and $B_d$ are $d$-dependent constants. [Note that we do not include here intermediate regimes, which leads to a slightly poorer quality of the approximation with respect to the $d=3$ case.]

Mathematically, the difference in the power law correction for a particle at the interface arises due to the different structure in the branchcut singularities in Eqs.~(\ref{potential},\ref{potentialddimensions}) with respect to the bulk case ($\kappa_1=\kappa_2$). Specifically, as $q\to i\kappa_2$ there is a divergence in the bulk case, but not at the interface. This is similar to what happens in polymer physics for a random or self-avoiding walk close to a surface~\cite{Duplantier1989}, where a similar change in the nature of the singularity leads to a change in the entropic exponent $\gamma$ (the power law correction), with no change in the connective constant (the leading exponential behaviour). For instance, the probability that a random walk with $N$ steps forms a loop in $d=1$ decays with $N$ as $N^{-1/2}$, but 
close to a hard surface the same probability is $\sim N^{-3/2}$~\cite{Maritan1999}.

Physically, these considerations suggest a mechanism for the change in asymptotic behaviour for the potential between two point charges at the interface, $U(r)$. The integrals determining this potential may be viewed as an integral of a propagator of a field, with $\kappa_{1,2}$ playing the role of the inverse mass, or as a correlator in Landau-Ginzburg theory~\cite{Chaikin1995}. Therefore the potential can be viewed as a sum of all contributions from interactions propagating from one particle to the other. The propagation occurs through lines which we call ``Debye strings''. Computing the potential then involves a summation over all fluctuating Debye strings. Because the two endpoints of a string are fixed at the interface (at the point charge positions), and because the screening in the first phase (i.e., water) is stronger, the strings are more likely to propagate through the second phase (i.e., oil). The statistics of the Debye strings contributing to the interaction is therefore different than in the bulk, where the propagation is symmetric, and the problem becomes qualitatively similar to that of a polymer close to a surface, thereby providing a simple physical picture to explain the change in the power law correction, or the anomalous asymptotic decay of the potential. 

In summary, we have computed the potential of a point charge at an interface between two electrolytes with distinct Debye length and dielectric permittivity, such as oil and water. Previous work focussed on the case where one of the Debye lengths is infinite -- so that there are no ions in one of the media -- and found a dipolar decay at long distances. Here, instead, we have seen that, when there are mobile charges in both electrolytes, the asymptotic behaviour of the potential is anomalous, and, quite notably, it differs from a screened charge monopole in the bulk and from a dipole at the interface. This result suggests that we should expect a non-trivial repulsive force between particles trapped at an oil-water interface, which is distinct from either a dipole-dipole or a simple screened Coulomb interaction. While a direct experimental test in self-assembled colloidal monolayers at an oil-water interface is possible, we anticipate it will be challenging, as previous experiments suggest that subtle charge redistribution effects, not captured by the standard Poisson-Boltzmann theory, may arise close to the interface~\cite{Vermant2010}.  Additionally, nonlinear effects unaccounted for by the Debye-H\"uckel theory may be important, as for air-water interfaces~\cite{Frydel2007}.

We expect that the anomalous scaling of the electrostatic potential will be relevant to understand the in-plane potential at salty interfaces~\cite{Netz1999} as well as protein-protein interactions in lipid bilayers. In these cases, the additional mobile charges at the interface can be dealt with, in the simplest instance, as a modified boundary conditions~\cite{Netz1999}. 
It will also be of interest to generalise our treatment to cases where the interface is curved, such as for a bijel, or for a droplet of water in oil. 

\newpage

\begin{widetext}

\section*{Appendix}
\subsection*{Interfacial potential for a 3-dimensional system: exact calculation}

In this Appendix we present the details of the whole calculation of the interparticle potential, of which we quoted the results in the main text.

 The interaction potential between two charged particles (each of charge $Q$) at a flat interface between two dielectric media -- for instance water and oil -- with dielectric permittivities $\epsilon_1$ and $\epsilon_2$ respectively, can be worked out by following~\cite{Stillinger1961,Hurd1985}, as done in the main text. The potential is given in terms of an integral involving special functions, as follows, 
\begin{equation}\label{potentialSI}
U(r)=\frac{Q^2}{2\pi r} \int_{0}^\infty dx \frac{x J_0(x)}{\epsilon_1 \sqrt{\kappa_1^2 r^2 + x^2} + \epsilon_2 \sqrt{\kappa_2^2 r^2 + x^2} }.
\end{equation}
where $J_0$ is the zero-th order Bessel function of the first kind. Note that $\kappa_1^{-1}$ and $\kappa_2^{-1}$ are the Debye length in the first and second phase. For a water/oil interface, typical parameters may be $\kappa_1\sim 10\kappa_2$ and $\epsilon_1\sim 40 \epsilon_2$ (these are relevant, for instance, to the case where the oil is dodecane~\cite{Muntz2019}).

We want to find the asymptotic behaviour ($r\to\infty$) of the potential. To do so, we need the asymptotic behaviour of
\begin{align}
I = \int_{0}^\infty dx \frac{x J_0(x)}{\epsilon_1 \sqrt{\kappa_1^2 r^2 + x^2} + \epsilon_2 \sqrt{\kappa_2^2 r^2 + x^2} }.
\end{align}

As mentioned in the main text, if we multiply the numerator and denominator in the integral in $I$ by $\epsilon_1 \sqrt{\kappa_1^2 r^2 + x^2} - \epsilon_2 \sqrt{\kappa_2^2 r^2 + x^2}$ we obtain
\begin{align}
I = \frac{1}{\epsilon_1^2 - \epsilon_2^2} \left( \epsilon_1 I_1 - \epsilon_2 I_2 \right),
\end{align}
where
\begin{align}
I_i =  \int_{0}^\infty dx \frac{x J_0(x) \sqrt{\kappa_i^2 r^2 + x^2}}{\alpha r^2 + x^2},
\end{align}
with $i=1,2$ and
\begin{align}
\alpha = \frac{\epsilon_1^2 \kappa_1^2 - \epsilon_2^2 \kappa_2^2  }{\epsilon_1^2 - \epsilon_2^2}.
\end{align}
Introducing $\delta_i \equiv \alpha - \kappa_i^2$, explicitly given by
\begin{align}
& \delta_1  =  \frac{\epsilon_{2}^2 (\kappa_1^2-\kappa_2^2)}{\epsilon_1^2 - \epsilon_2^2} \\ \nonumber
& \delta_2 =  \frac{\epsilon_{1}^2 (\kappa_1^2-\kappa_2^2)}{\epsilon_1^2 - \epsilon_2^2},
\end{align}
and Taylor-expanding $I_i$ with respect to $\delta_i$, we obtain
\begin{align}
I_i = \sum_{m=0}^\infty (-1)^m (\delta_i r^2)^m  \int_{0}^\infty dx \frac{x J_0(x) }{\left(\kappa_i^2 r^2 + x^2\right)^{m+1/2}}.
\end{align}
These integrals can be performed as discussed in~\cite{Hurd1985} to give
\begin{align}
I_i = \sum_{m=0}^\infty (-1)^m (\delta_i r^2)^m  \frac{2^{1/2-m}}{\Gamma(m+1/2)} \frac{K_{m-1/2}(\kappa_i r)}{(\kappa_i r)^{m-1/2}}
= \sqrt{2 \kappa_i r} \sum_{m=0}^\infty (-1)^m\left( \frac{\delta_i}{4\kappa_i^2}2\kappa_i r \right)^m \frac{K_{m-1/2}(\kappa_i r)}{\Gamma(m+1/2)},
\end{align}
where $\Gamma$ denotes the $\Gamma$ function~\cite{Abramowitz1972}.
Now we separate the term with $m=0$ (which gives a screened monopole behaviour) and shift the index from $m$ to $\mu + 1$, to get
\begin{align}
I_i = \sqrt{2 \kappa_i r}\Bigg[ \frac{K_{-1/2}(\kappa_i r)}{\sqrt{\pi}} - 
\sum_{\mu=0}^\infty (-1)^\mu\left( \frac{\delta_i}{4\kappa_i^2}2\kappa_i r \right)^{\mu+1} \frac{K_{\mu+1/2}(\kappa_i r)}{\Gamma(\mu+3/2)}
\Bigg].
\end{align}
We next use the series representation of the half-integer modified Bessel function of the second kind~\cite{Abramowitz1972}, 
\begin{align}
K_{\mu+1/2}(z) = e^{-z} \sqrt{\frac{\pi}{2z}}\sum_{p=0}^\mu \frac{(\mu+p)!}{p! (\mu-p)!}\frac{1}{(2z)^p}.
\end{align}
This gives
\begin{align}
I_i = \sqrt{2 \kappa_i r}\Bigg[ \frac{K_{-1/2}(\kappa_i r)}{\sqrt{\pi}} - 
e^{-\kappa_i r}\sqrt{\frac{\pi}{2\kappa_i r}}
\sum_{\mu=0}^\infty \sum_{p=0}^{\mu} (-1)^\mu\left( \frac{\delta_i}{4\kappa_i^2} \right)^{\mu+1} (2\kappa_i r)^{\mu+1-p} \frac{(\mu+p)!}{p! (\mu-p)!}\
\frac{1}{\Gamma(\mu+3/2)}
\Bigg].
\end{align}
Now, we change the order of summation in the two series, 
\begin{align}
I_i = \sqrt{2 \kappa_i r}\Bigg[ \frac{K_{-1/2}(\kappa_i r)}{\sqrt{\pi}} - 
e^{-\kappa_i r}\sqrt{\frac{\pi}{2\kappa_i r}}
\sum_{p=0}^\infty \sum_{\mu=p}^{\infty} (-1)^\mu\left( \frac{\delta_i}{4\kappa_i^2} \right)^{\mu+1} (2\kappa_i r)^{\mu+1-p} \frac{(\mu+p)!}{p! (\mu-p)!}\
\frac{1}{\Gamma(\mu+3/2)}
\Bigg].
\end{align}
The inner series can be re-summed, to finally get
\begin{align}
I_i = e^{-\kappa_i r} - e^{-\kappa_i r} \frac{r\delta_i}{\kappa_i} \sum_{p=0}^{\infty} \frac{(-1)^p}{2p+1} \left( \frac{\delta_i}{\kappa_i^2}\right)^p 
{}_1F_1\left(2p+1;\frac{3}{2}+p;-\frac{r\delta_i}{2\kappa_i} \right),
\end{align}
where ${}_1F_1$ is the confluent hypergeometric function of the first kind~\cite{Abramowitz1972}. 

Its asymptotic behaviour (for large $\frac{r\delta_i}{2\kappa_i}\equiv z$) is given by~\cite{Abramowitz1972}
\begin{align}
{}_1F_1(a;b;-z) \sim \frac{\Gamma(b)}{\Gamma(a)} e^{-z} (-z)^{a-b} \left[1-\frac{(a-1)(a-b)}{z}+\frac{(a-2)(a-1)(a-b-1)(a-b)}{2z^2}+\cdots\right]+ \\ \nonumber
\frac{\Gamma(b)}{\Gamma(b-a)}\frac{1}{z^a}\left[1+\frac{a(a-b+1)}{z}+\frac{a(a+1)(a-b+1)(a-b+2)}{2z^2}+\cdots\right].
\end{align}
The first sum is subdominant, if the second one exists, and for our combination of indices, it does. Therefore, we obtain the following asymptotic series (valid for $r\to \infty$),
\begin{align}\label{asymptotic0SI}
& I_i\sim e^{-\kappa_i r}- e^{-\kappa_i r}\left[1+\frac{\kappa_i}{r\delta}+\frac{3\kappa_i^2}{r^2\delta^2}+\frac{1}{r^2\delta}+{\mathcal O}\left(\frac{1}{r^3}\right)\right]= -e^{-\kappa_i r}\left[\frac{\kappa_i}{r\delta}+\frac{3\kappa_i^2}{r^2\delta^2}+\frac{1}{r^2\delta}+{\mathcal O}\left(\frac{1}{r^3}\right)\right].
\end{align}
Consequently, as the integral $I$ is dominated by the behaviour of $I_2$ for $r\to\infty$ (as $\kappa_2<\kappa_1$ for a water/oil interface), we obtain that
\begin{align}\label{asymptotic1}
I = \frac{1}{\epsilon_1^2 - \epsilon_2^2} \left( \epsilon_1 I_1 - \epsilon_2 I_2 \right) \sim e^{-\kappa_2 r}\left[\frac{\epsilon_2 \kappa_2}{\epsilon_1^2(\kappa_1^2-\kappa_2^2)r}+\frac{3\epsilon_2(\epsilon_1^2-\epsilon_2^2)\kappa_2^2}{\epsilon_1^4(\kappa_1^2-\kappa_2^2)^2r^2}+\frac{\epsilon_2}{\epsilon_1^2(\kappa_1^2-\kappa_2^2)r^2}\right].
\end{align}
The numerical solution of $I$ is compared to this approximation in Fig.~\ref{fig1SI}.

\begin{figure}[!h]
\centerline{\includegraphics[width=0.75\textwidth, angle=0]{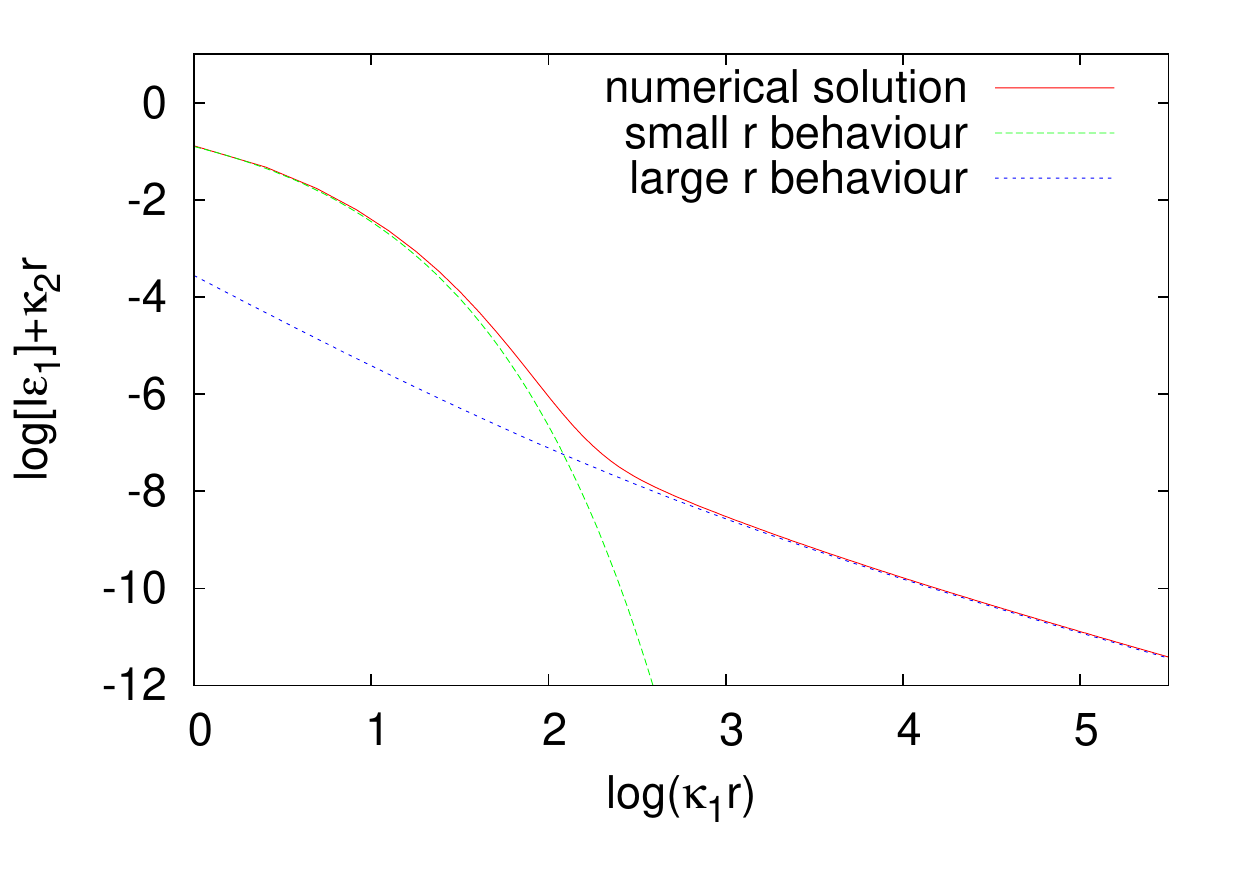}}
\caption{Plot of the numerical solution of $\log(I\epsilon_1 e^{\kappa_2 r})$ versus $\log(\kappa_1 r)$, showing the crossover between small $r$ behaviour, corresponding to a screened monopole with decay constant $\kappa_1$, and a large $r$ behaviour, described by Eq.~\ref{asymptotic1}. Parameters are: $\kappa_1=10\kappa_2$, $\epsilon_1=40\epsilon_2$, relevant for a water-oil interface.}
\label{fig1SI}
\end{figure}

The corresponding asymptotic leading behaviour for the interaction potential $U$ is 
\begin{align}\label{asymptotic2SI}
U(r) \sim \frac{Q^2}{2\pi}\frac{\epsilon_2\kappa_2}{\epsilon_1^2(\kappa_1^2-\kappa_2^2)}\frac{e^{-\kappa_2r}}{r^2},
\end{align}
which decays faster than a screened monopole with decay constant $\kappa_2$.

\begin{figure}[!h]
\centerline{\includegraphics[width=0.75\textwidth, angle=0]{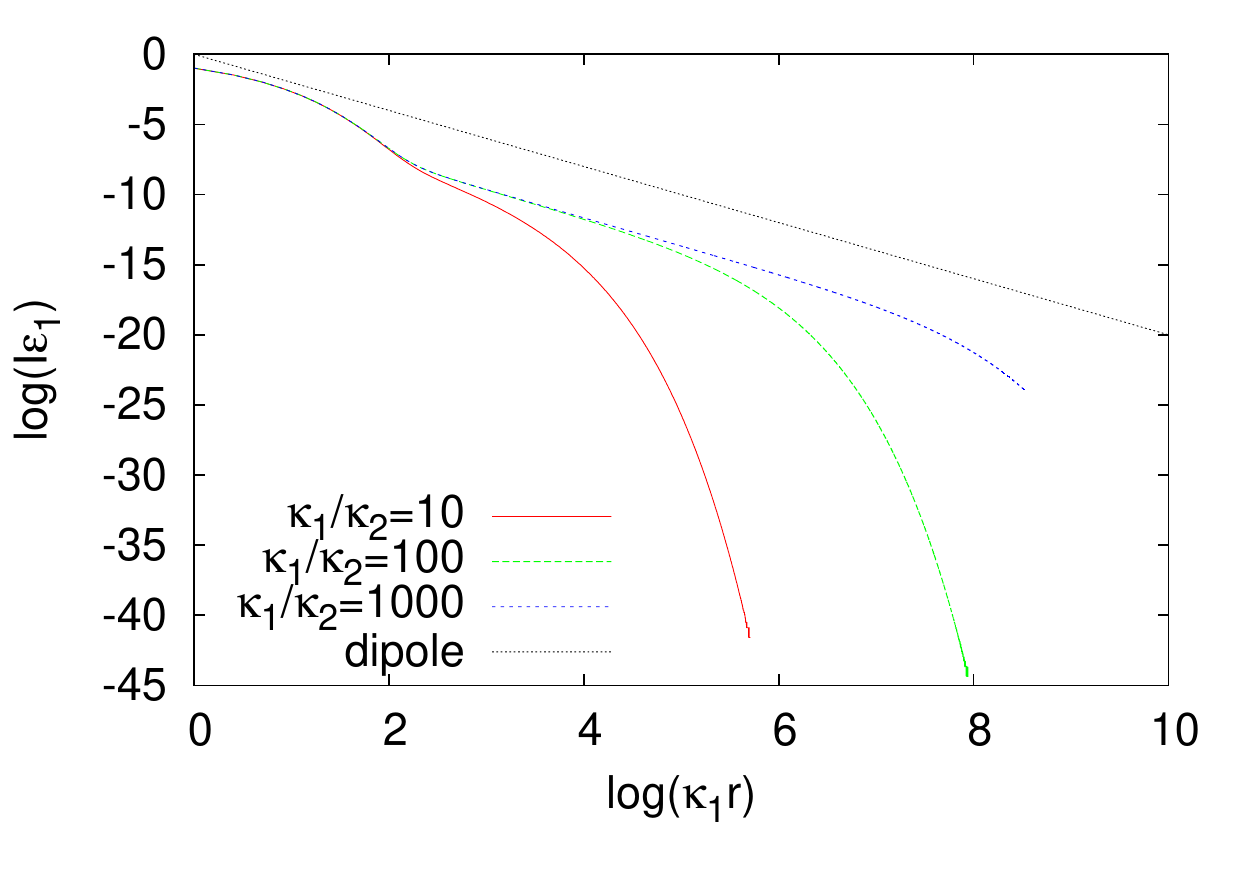}}
\caption{Plot of the numerical solution of $\log(I\epsilon_1)$ versus $\log(\kappa_1 r)$, for different values of $\kappa_2$, showing an intermediate dipole-like decay only for very small $\kappa_2$. Curves are calculated with $\epsilon_1=40\epsilon_2$.}
\label{dipoleoilwater}
\end{figure}

For a generic oil-water interface, we can consider that $\epsilon_2\ll \epsilon_1$ and that $\kappa_2\ll \kappa_1$. In this limit, as done in the text, it is useful to write the potential as a sum of three contribution, a screened monopole one, a ``screened dipole'' term $\sim \frac{e^{-\kappa_2 r}}{r^3}$, and a term $\frac{e^{-\kappa_2 r}}{r^2}$, as follows,
\begin{equation}\label{fitSI}
U(r) \simeq \frac{Q^2}{2\pi \epsilon_1} \frac{e^{-\kappa_1 r}}{r} + 
\frac{Q^2\epsilon_2}{2\pi \epsilon_1^2 \kappa_1^2}\frac{e^{-\kappa_2 r}}{r^3}+\frac{Q^2\epsilon_2 \kappa_2}{2\pi \epsilon_1^2 \kappa_1^2}\frac{e^{-\kappa_2 r}}{r^2}.
\end{equation}
The screened monopole contribution dominates at small $r$, whereas the $\frac{e^{-\kappa_2 r}}{r^2}$ is the asymptotic contribution, which dominates for $r\to\infty$. The screened dipole term $\sim \frac{e^{-\kappa_2 r}}{r^3}$ is the one which gives the dipolar interaction if $\kappa_2=0$. For finite $\kappa_2$, it can be seen at intermediate values of $r$, but only under some conditions. Taking $\epsilon_1=40\epsilon_2$, Fig.~\ref{dipoleoilwater} shows that an apparent dipolar contribution can only be seen if $\kappa_1 \gg 10 \kappa_2$.

\subsection*{Interfacial potential for a $d$-dimensional system: asymptotic behaviour}

We now consider the case of a $d$-dimensional system (and a $(d-1)$-dimensional interface). The potential can be obtained by following the same procedure used in $d=3$ to be,
\begin{eqnarray}\label{potentialddimensionsSI}
\phi(r) & = & \frac{Q}{\left(2\pi\right)^{d-1}} \int d^{d-1} {\mathbf q} \frac{e^{i {\mathbf q}\cdot \rr}}{\epsilon_1 \sqrt{\kappa_1^2 + q^2} + \epsilon_2 \sqrt{\kappa_2^2 + q^2} } \\ \nonumber 
& = & \frac{Q}{\left(2\pi\right)^{\frac{d-1}{2}} r^{d-2}} \int_{0}^\infty dx \frac{x^{\frac{d-1}{2}} J_{\frac{d-3}{2}}(x)}{\epsilon_1 \sqrt{\kappa_1^2 r^2 + x^2} + \epsilon_2 \sqrt{\kappa_2^2 r^2 + x^2} }.
\end{eqnarray}
To see that the two expressions in Eq.~(\ref{potentialddimensionsSI}) are equivalent, it is useful to note the following identity
\begin{equation}
\int_{0}^{\pi} \frac{d\theta}{2\pi}  e^{iq \cos{\theta}} \left(\sin{\theta}\right)^n  = 2^{\frac{n}{2}}\sqrt{\pi}J_{\frac{n}{2}}(q)\Gamma(\frac{n+1}{2}),
\end{equation}
where $n\ge 0$ is a non-negative integer.

We now study separately the case where $\kappa_1=\kappa_2$ (no interface, or bulk behaviour), and that of $\kappa_1\ne \kappa_2$, with both $\kappa_1$ and $\kappa_2$ non-zero.

\subsubsection*{$\kappa_1=\kappa_2$ case: the bulk Yukawa potential case}

We start by analysing the behaviour of the following $d$-dimensional integral,
\begin{equation}
C_d = \frac{1}{(2\pi)^{d-1}} \int d^{d-2} {\mathbf q_{\perp}} \int_{-\infty}^{\infty} dq_{\parallel} \frac{e^{i q_{\parallel}r}}{\sqrt{\kappa^2 + q_{\perp}^2+q_{\parallel}^2}},
\end{equation}
which equals the interparticle potential in the bulk up to a multiplicative constant (equal to $\frac{Q^2}{2\epsilon}$). While this is well known, it is useful to solve this integral in a way which can be generalised to the interface case.

Introducing the $(d-3)$-dimensional unit sphere (i.e., the surface of a unit sphere embedded in a $(d-2)$-dimensional space), $S_{d-3}=\frac{2\pi^{\frac{d-2}{2}}}{\Gamma(\frac{d-2}{2})}$, and performing a ``Wick rotation'' in the $q_{\parallel}$ integral, equivalent to sending $q_{\parallel}\to iq_{\parallel}$, we obtain
\begin{eqnarray}
C_d & = &  \frac{S_{d-3}}{(2\pi)^{d-2}}\int_0^{+\infty} dq_{\perp} q_{\perp}^{d-3} \int_{-\infty}^{\infty} \frac{dq_{\parallel}}{2\pi} \frac{e^{i q_{\parallel}r}}{\sqrt{\kappa^2 + q_{\perp}^2+q_{\parallel}^2}} \\ \nonumber
 & = & \frac{S_{d-3}}{(2\pi)^{d-2}}\int_0^{+\infty} dq_{\perp} q_{\perp}^{d-3} \int_{+i \infty}^{-i \infty} \frac{i dq_{\parallel}}{2\pi} \frac{e^{-q_{\parallel}r}}{\sqrt{\kappa^2 + q_{\perp}^2-q_{\parallel}^2}} \\ \nonumber
 & = & \frac{S_{d-3}}{(2\pi)^{d-2}}\int_0^{+\infty} dq_{\perp} q_{\perp}^{d-3} \int_{-i \infty}^{+i \infty} \frac{dq_{\parallel}}{2\pi i} \frac{e^{q_{\parallel}r}}{\sqrt{\kappa^2 + q_{\perp}^2-q_{\parallel}^2}}.  
\end{eqnarray}
The integral over $q_{\parallel}$ is now equivalent to an inverse Laplace transform, whose asymptotic behaviour for $r\to \infty$ is dominated by its singularities, here a branchpoint at $q=-q_0\equiv -\sqrt{q_{\perp}^2+\kappa^2}$ (chosen so as to be able to close the Bromwich integration contour in the integral over $q_{\parallel}$ with paths in the complex plane over which the integral exists). Therefore, for $r\to \infty$, we obtain
\begin{equation}
C_d \simeq \frac{S_{d-3}}{(2\pi)^{d-2}}\int_0^{+\infty} dq_{\perp} q_{\perp}^{d-3} \int_{-q_0-i \infty}^{-q_0+i \infty} \frac{dq_{\parallel}}{2\pi i} e^{-q_0 r}\frac{e^{q_{\parallel}r}} {\sqrt{2q_0}\sqrt{q_0+q_{\parallel}}} 
\end{equation}
By using the inverse Laplace transform identity
\begin{equation}
\int_{-i \infty}^{+i \infty} \frac{dq_{\parallel}}{2\pi i}\frac{e^{q_{\parallel}r}} {\sqrt{q}} = \frac{1}{\Gamma(1/2)r^{1/2}}=\frac{1}{\sqrt{\pi}r^{1/2}}
\end{equation}
we get
\begin{eqnarray}\label{qperpintegralbulk}
C_d & \simeq & \frac{S_{d-3}}{(2\pi)^{d-3/2}r^{1/2}}\int_0^{+\infty} dq_{\perp} q_{\perp}^{d-3} \frac{e^{-rq_0}}{\sqrt{q_0}} \\ \nonumber
& \simeq & \frac{S_{d-3}}{(2\pi)^{d-3/2}r^{1/2}} \int_{0}^{+\infty} dq_{\perp} q_{\perp}^{d-3} \frac{e^{-r\kappa}e^{-\frac{rq_{\perp}^2}{2\kappa}}}{\sqrt{\kappa}},
\end{eqnarray}
where in the last step we have approximated the integrand for $q_{\perp}\to 0$, as this will give the dominant contribution in the $r\to \infty$ limit. [Equivalently, we could have performed a saddle point approximation here.]

By rescaling $q_{\perp}$ and performing the final integral, we get
\begin{eqnarray}\label{yukawaasymptotics}
C_d & \simeq & \frac{S_{d-3}2^{\frac{d-2}{2}}}{(2\pi)^{d-3/2}} \frac{{\kappa}^{\frac{d-3}{2}}e^{-\kappa r}}{r^{\frac{d-1}{2}}} \int_{0}^{+\infty} dq_{\perp} q_{\perp}^{d-3} e^{-q_{\perp}^2} \\ \nonumber
& = & \frac{2^{\frac{d}{2}-1}S_{d-3}\Gamma(\frac{d}{2}-1)}{2(2\pi)^{d-3/2}} \frac{{\kappa}^{\frac{d-3}{2}}e^{-\kappa r}}{r^{\frac{d-1}{2}}} \\ \nonumber
& = & \frac{1}{(2\pi)^{\frac{d-1}{2}}} \frac{{\kappa}^{\frac{d-3}{2}}e^{-\kappa r}}{r^{\frac{d-1}{2}}}.
\end{eqnarray}
The resulting scaling is the well-known scaling of the $d$-dimensional Yukawa potential, or of the Gaussian correlator in the $d$-dimensional Landau-Ginzburg theory for the critical transition in a magnet~\cite{Chaikin1995}.


\subsubsection*{$\kappa_1\ne\kappa_2$ case: asymptotics for the interfacial potential}

We now consider the behaviour of the interfacial potential. For simplicity we set $\epsilon_1=\epsilon_2=1$ (the asymptotic behaviour is the same as for $\epsilon_1\ne \epsilon_2$). The potential is proportional to the following integral,
\begin{eqnarray}
C_d & = &  \frac{S_{d-3}}{(2\pi)^{d-2}}\int_0^{+\infty} dq_{\perp} q_{\perp}^{d-3} \int_{-\infty}^{\infty} \frac{dq_{\parallel}}{2\pi} \frac{e^{i q_{\parallel}r}}{\sqrt{\kappa_1^2 + q_{\perp}^2+q_{\parallel}^2}+\sqrt{\kappa_2^2 + q_{\perp}^2+q_{\parallel}^2}} \\ \nonumber
 & = & \frac{S_{d-3}}{(2\pi)^{d-2}}\int_0^{+\infty} dq_{\perp} q_{\perp}^{d-3} \int_{-\infty}^{\infty} \frac{dq_{\parallel}}{2\pi} \frac{e^{i q_{\parallel}r}}{\kappa_1^2-\kappa_2^2}\left[{\sqrt{\kappa_1^2 + q_{\perp}^2+q_{\parallel}^2}-\sqrt{\kappa_2^2 + q_{\perp}^2+q_{\parallel}^2}}\right] \\ \nonumber 
& \equiv & C_{d,1}+C_{d,2}.  
\end{eqnarray}
As the second integral, $C_{d,2}$, describes propagation of the interaction in the second medium, and as $\kappa_2<\kappa_1$, this will be the dominant contribution at large $r$. Performing a Wick rotation in the integral over $q_{\parallel}$, and using the following inverse Laplace transform identity (which involves regularisation of the integral),
\begin{equation}
\int_{-i \infty}^{+i \infty} \frac{dq_{\parallel}}{2\pi i} e^{q_{\parallel}r} \sqrt{q} = \frac{1}{\Gamma(-1/2)r^{3/2}}=-\frac{1}{2\sqrt{\pi}r^{3/2}},
\end{equation}
we obtain
\begin{eqnarray}
C_{d,2} & = & -\frac{S_{d-3}}{(2\pi)^{d-2}(\kappa_1^2-\kappa_2^2)}\int_0^{+\infty} dq_{\perp} q_{\perp}^{d-3} \int_{-\infty}^{\infty} \frac{dq_{\parallel}}{2\pi} e^{i q_{\parallel}r} \sqrt{\kappa_2^2 + q_{\perp}^2+q_{\parallel}^2} \\ \nonumber
& = & -\frac{S_{d-3}}{(2\pi)^{d-2}(\kappa_1^2-\kappa_2^2)}\int_0^{+\infty} dq_{\perp} q_{\perp}^{d-3} \int_{-i \infty}^{+i \infty} \frac{dq_{\parallel}}{2\pi i} e^{q_{\parallel}r} \sqrt{\kappa_2^2 + q_{\perp}^2-q_{\parallel}^2} \\ \nonumber
& \simeq & -\frac{S_{d-3}}{(2\pi)^{d-2}(\kappa_1^2-\kappa_2^2)}\int_0^{+\infty} dq_{\perp} q_{\perp}^{d-3} \int_{-i \infty}^{+i \infty} \frac{dq_{\parallel}}{2\pi i} e^{q_{\parallel}r} \sqrt{2q_0} \sqrt{q_0+q_{\parallel}} \\ \nonumber
& = & \frac{S_{d-3}}{(2\pi)^{d-2}(\kappa_1^2-\kappa_2^2)} \frac{1}{2\sqrt{\pi} r^{3/2}}\int_0^{+\infty} dq_{\perp} q_{\perp}^{d-3} e^{-rq_0} \sqrt{2q_0} \\ \nonumber
q_0 & = & \sqrt{q_{\perp}^2+\kappa_2^2}.
\end{eqnarray}
As in Eq.~(\ref{qperpintegralbulk}), we now approximate the integral over $q_{\perp}$ by noting that, for large $r$, it is dominated by the behaviour for $q_{\perp}\to 0$, to obtain
\begin{eqnarray}\label{interfacialasymptotics}
C_{d,2} & \simeq & \frac{S_{d-3}}{(2\pi)^{d-3/2}}\frac{\sqrt{\kappa_2}}{(\kappa_1^2-\kappa_2^2)} \frac{e^{-\kappa_2 r}}{r^{3/2}}\int_0^{+\infty} dq_{\perp} q_{\perp}^{d-3} e^{-r\frac{q_{\perp}^2}{2\kappa_2}} \\ \nonumber
& = & \frac{1}{(2\pi)^{\frac{d-1}{2}}} \frac{\kappa_2^{\frac{d-1}{2}}e^{-\kappa_2 r}}{(\kappa_1^2-\kappa_2^2)r^{\frac{d+1}{2}}}.
\end{eqnarray}
The resulting scaling coincides with that worked out previously for the special case of $d=3$. It can be seen that for any $d\ge 3$ the limits $\kappa_2\to 0$ and $\kappa_2\to \kappa_1$ are both singular. The former limit gives zero, which means that a higher-order decay in $r$ becomes relevant, and indeed $C_{d,2}\sim 1/r^{d-2}$ for $\kappa_2\to 0$. The latter gives infinity, which means that the potential decays more slowly than in Eq.~(\ref{interfacialasymptotics}), and indeed in that limit we obtain the bulk Yukawa potential in Eq.~(\ref{yukawaasymptotics}).

To approximate the interparticle interaction potential at the interface, $U(r)$, for cases in which $\epsilon_2\ll \epsilon_1$ (as in oil-water interfaces), we can use a similar argument as done in the $d=3$ case (see main text), but including only the small $r$ and large $r$ behaviours. We therefore fit the data with a sum of a Yukawa-like potential with decay constant $\kappa_1$, which should describe the behaviour at small distances, and the asymptotic behaviour in Eq.~(\ref{interfacialasymptotics}) at large distances,
\begin{equation}\label{fitddimensionsSI}
\frac{U(r)\epsilon_1}{Q^2} = A_d \frac{e^{-\kappa_1 r}}{r^{\frac{d-1}{2}}} + B_d \frac{e^{-\kappa_2 r}}{r^{\frac{d+1}{2}}}.
\end{equation}
The constants $A_d$ and $B_d$ can be set according to Eq.~(\ref{yukawaasymptotics}) and Eq.~(\ref{interfacialasymptotics}) respectively, or fitted as free parameters. Figure~\ref{ddimfit} shows the comparison between the numerical solutions for the interparticle potential in $d=4, 5$ and the approximations (or fits) in Eq.~(\ref{fitddimensionsSI}). The asymptotic behaviour is captured well by Eq.~(\ref{interfacialasymptotics}); the Yukawa-like decay provides a poorer approximation at low distances, which is reasonable as in general in $d\ne 3$ Eq.~(\ref{yukawaasymptotics}) is not exact, but is itself only asymptotically correct. Allowing $A_d$ and $B_d$ to be free parameters leads to a reasonable approximation for the numerical solution over the whole range of distances which we calculated.

\begin{figure}[!h]
\subfloat[]{\includegraphics[width = .44\textwidth]{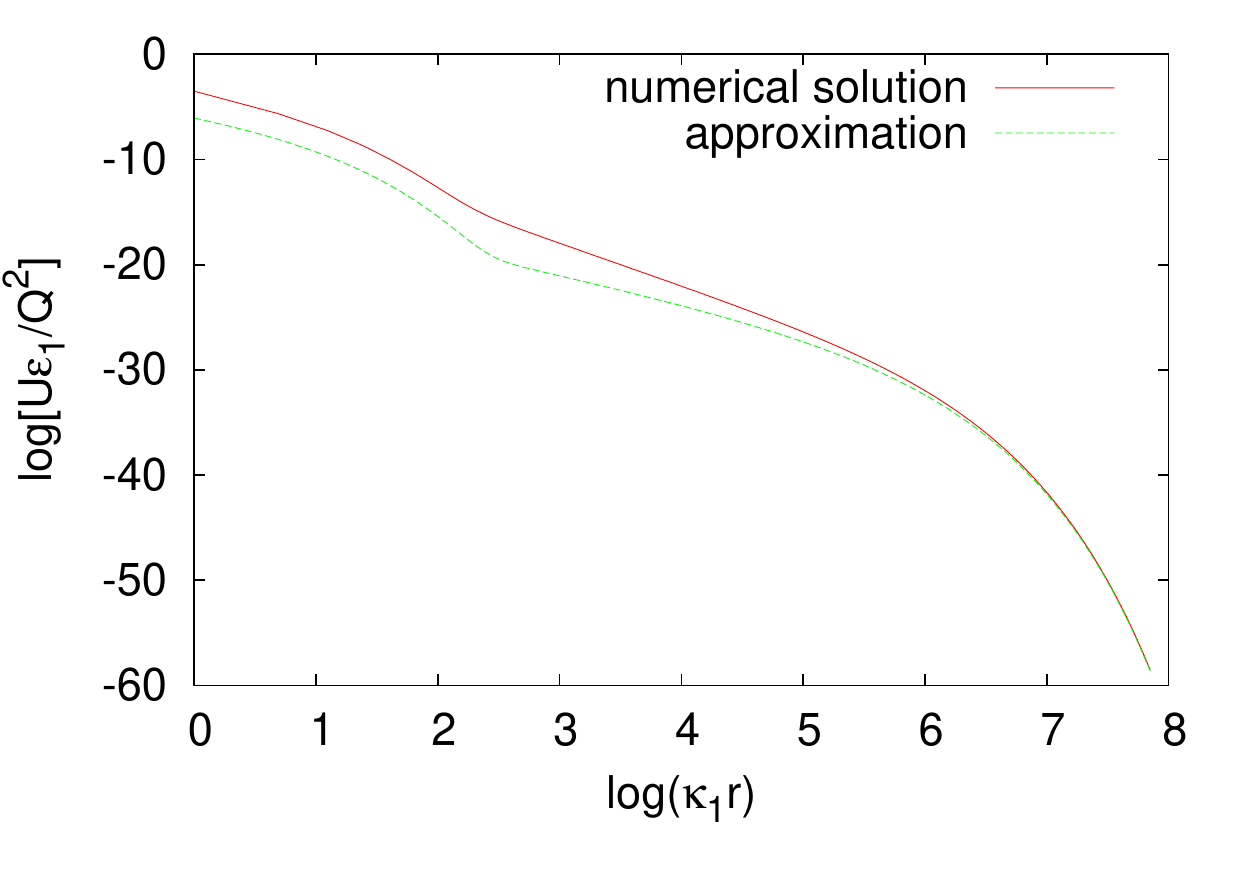}}
\renewcommand{\thesubfigure}{b}
\subfloat[]{\includegraphics[width = .44\textwidth]{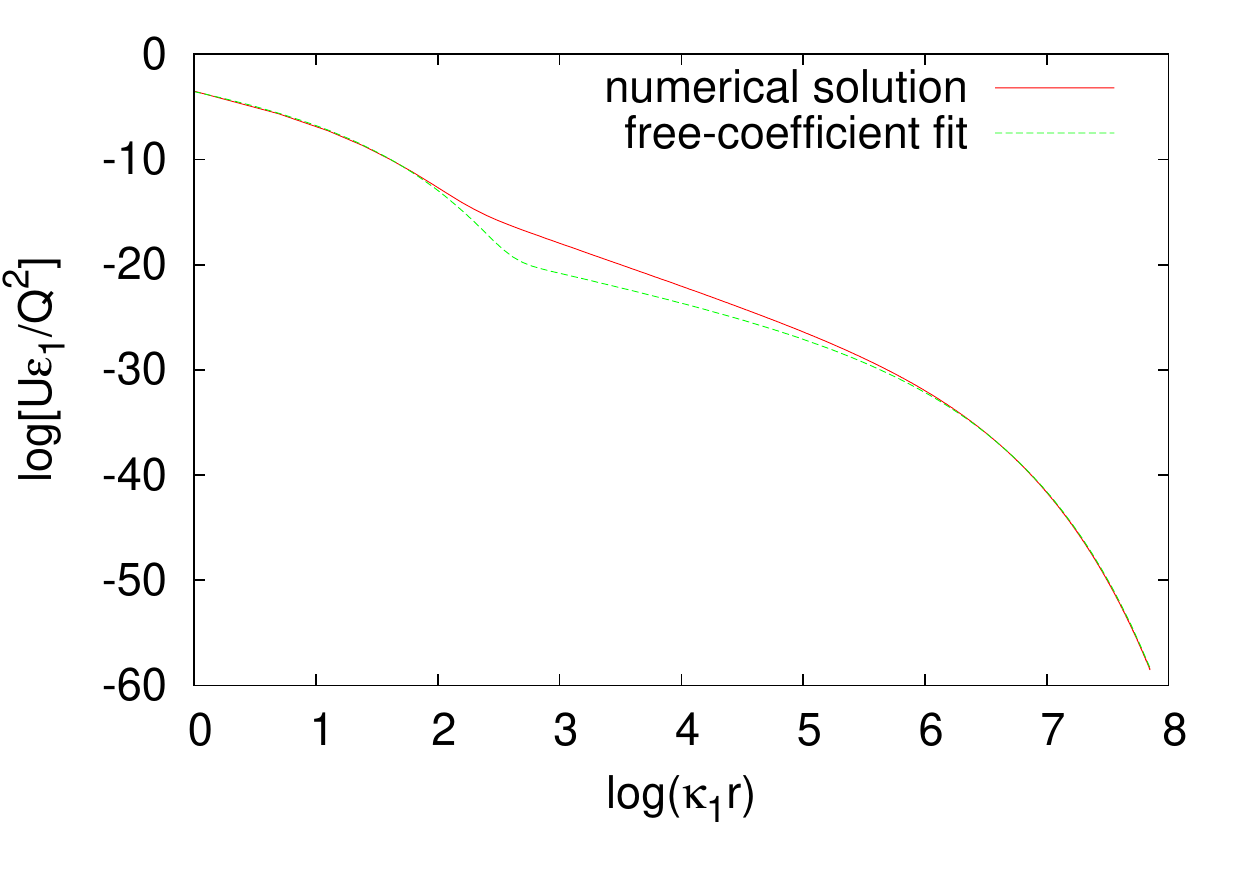}}
\renewcommand{\thesubfigure}{c}
\subfloat[]{\includegraphics[width = .44\textwidth]{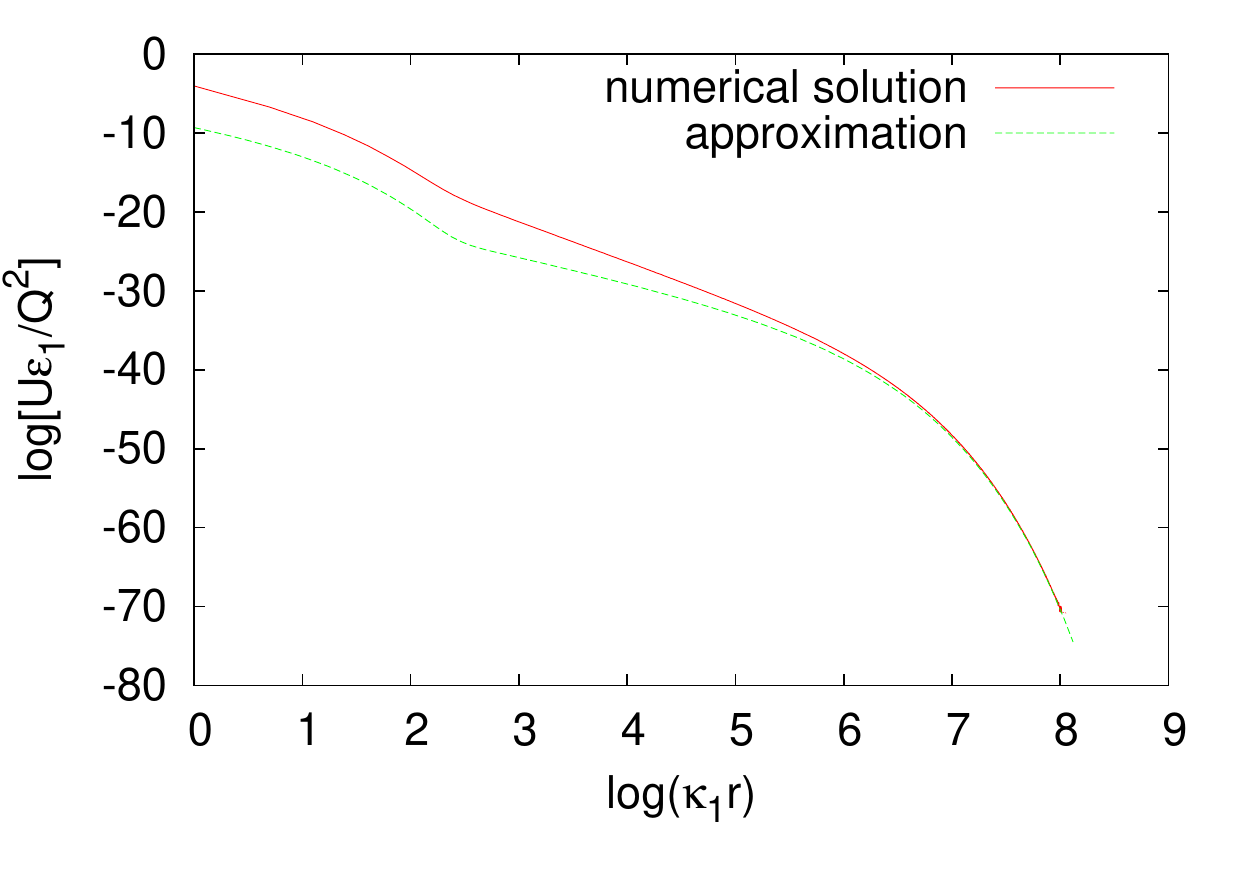}}
\renewcommand{\thesubfigure}{d}
\subfloat[]{\includegraphics[width = .44\textwidth]{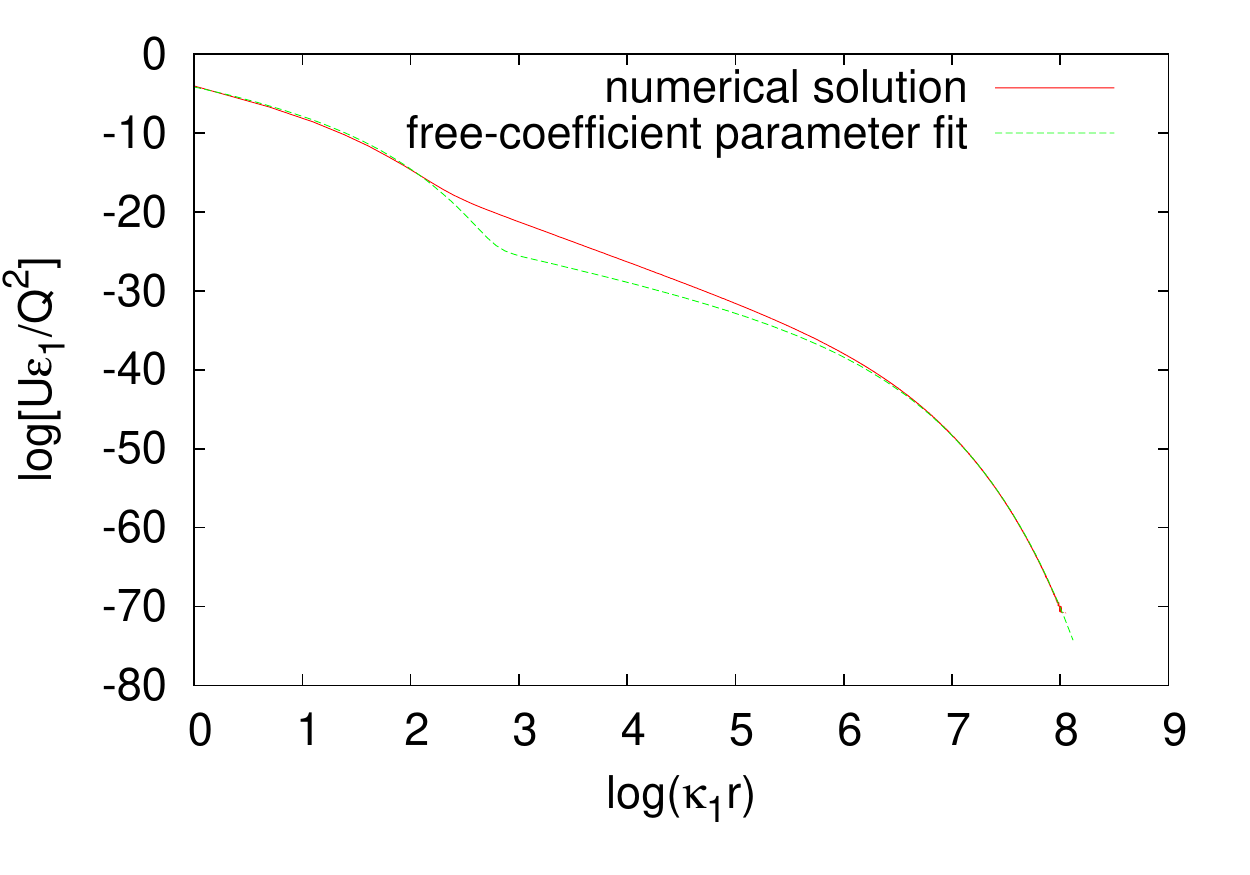}}
\caption{(a) Comparison between the numerical solution for $U(r)\epsilon_1/Q^2$ versus $r$ in $d=4$ and Eq.~(\ref{fitddimensionsSI}), with constants chosen according to the asymptotic analysis. (b) Comparison to a fit where $A_4$ $B_4$ were separately fitted to the low and high distances regimes respectively. The fit corresponds to Eq.~(\ref{fitddimensionsSI}) with $A_4\simeq 0.0779$ and $B_4\simeq 2.02\times 10^{-6}$. (c) Same as (a), but for $d=5$. (d) Same as (b), but for $d=5$.  The fit corresponds to Eq.~(\ref{fitddimensionsSI}) with $A_5\simeq 0.0434$ and $B_5\simeq 7.90\times 10^{-8}$.  }
\label{ddimfit}
\end{figure}

\subsection*{Numerical details}

To evaluate the interparticle potential $U(r)$ (or the potential $\phi(r)$), we need to calculate the quantity $I_d$ in Eq.~(12) in the main text (equivalently the second row of Eq.~(\ref{potentialddimensionsSI}) in the Appendix).

As the integrands are highly oscillating functions, care needs to be taken in this numerical evaluation. For our purposes, we used the ``quadosc'' quadrature function in python. For $d=3$, we split up the integration range between the zeroes of $J_0(x)$; for $d>3$, we split up the range uniformly (with a period of $2\pi$). [In $d=3$, the methods are equivalent for the precision which we require.]

\end{widetext}

\end{document}